## Закономерности институционализации электронной коммерции
*Калужский М.Л.*

***Аннотация***: *Статья об объективных закономерностях формирования социально-экономических институтов в системе электронной коммерции. Бурное развитие интернет-технологий стало причиной глубокой институциональной трансформации экономических отношений. Автор анализирует значение трансакционных издержек как движущей силы формирования новых экономических институтов в сетевой экономике.*

## Laws governing development of institutionalization of e-commerce
*Kaluzhsky **M.L.**.*

***Abstract***: *Article about objective laws of formation of social and economic institutes in system of electronic commerce. Rapid development of Internet technologies became the reason of deep institutional transformation of economic relations. The author analyzes value transaction costs as motive power of formation of new economic institutes in network economy.*

***Ключевые слова***: *сетевая экономика, институционализм, интернет, трансакционные издержки, электронная коммерция, интернет-коммуникации, институциональная теория, интернет-маркетинг, жизненный цикл.*

Институционализация электронной коммерции обусловлена особенностями её происхождения и дальнейшего развития. Возраст электронной коммерции по различным оценкам составляет около полувека [1, с. 15]. За это непродолжительное время произошла качественная трансформация электронной коммерции.

Первоначально электронная коммерция представляла собой форму торговой деятельности, являясь одним из инструментов продаж. Она воспроизводила методы традиционной торговли, перенося их в виртуальную среду. Не случайно некоторые авторы в России до сих пор рассматривают Интернет лишь в качестве среды «*для построения коммуникаций*» [2, с. 12].

Однако на рубеже тысячелетий электронная коммерция превратилась в самостоятельный вид торговой деятельности так же, как Интернет превратился из коммуникативной среды в динамично развивающийся глобальный рынок [3, с. 4]. Это неизбежно привело к формированию особых, присущих только электронной коммерции, институтов и институциональных отношений.

**Институциональные процессы в электронной коммерции**. Процессы формирования институтов электронной коммерции пока ещё не нашли должного отражения в институциональной теории. Положение несколько усугубляется тем, что далеко не все постулаты институциональной (и неоклассической) теории находят своё подтверждение в повседневной практике сетевой экономики, исходя из её особенностей.

Особо следует отметить, что электронная коммерция имеет дело не реальными товарами, а с информацией об этих товарах и связанных с ними трансакциями. Продавцы, посредники и покупатели в виртуальной среде Интернета извлекают выгоду не от сделок по продаже товаров, а от оказания информационных услуг, связанных с заключением таких сделок. Это принципиальное отличие электронной коммерции от коммерции традиционной.

Кстати, несмотря на то, что экономическая теория традиционно не рассматривает информацию в качестве самостоятельного фактора экономической деятельности, именно неоинституциональная теория уделяет ему наибольшее внимание. Так, например, один из столпов неоинституционализма К.Эрроу прямо указывает, что «*информация – это экономическая категория, т.е. товар, имеющий стоимость*» и что «*… экономическая роль информации заключается в снижении неопределенности и предотвращении убытков*» [4, с. 100].

Вместе с тем, именно электронная коммерция является сферой деятельности, в наибольшей степени связанной с экономическим обменом информацией. Можно даже сказать, что электронная коммерция восполняет неполноту использования потенциала информации как товара в экономике. Особенность электронной коммерции заключается в том, что в качестве товара здесь выступает даже не информация как таковая, а электронные коммуникации, многократно повышающие экономический эффект от использования информации.

М.Кастельс писал по этому поводу: «*Сущность электронного бизнеса заключается в обеспечиваемой Интернетом интерактивной сетевой связи между производителями, потребителями и поставщиками услуг*» [5, с. 96]. Конкурентоспособность участников рынка электронной коммерции напрямую зависит от степени их информированности. Поэтому сокращение дисбаланса между действительными и потенциальными возможностями информационных коммуникаций привело к ускоренному развитию электронной коммерции.

Определяющую роль здесь сыграли специфические особенности электронных трансакций. В традиционной экономике сложился некий баланс, отражающий структуру трансакционных издержек. Этот баланс был основан на объективной невозможности сокращения издержек в рамках существующих экономических и институциональных отношений.

По определению Д.Норта: «*Трансакционные издержки возникают вследствие того, что передача и получение информации сопровождаются издержками, что … любые усилия "актеров" по структурированию взаимоотношений с другими людьми с помощью институтов приводят к той или иной степени несовершенства рынков*» [6, с. 138-139]. Иначе говоря, традиционная структура трансакционных издержек была затратна, но в рамках существующих институциональных отношений улучшить её было невозможно.

Электронная коммерция открыла новые институциональные возможности изменения структуры трансакционных издержек. «*Необходимость переработки информации вследствие затратного характера трансакций,* – отмечает Д.Норт, – *лежит в основе образования институтов*» [6, с. 138]. Процесс институционализации электронной торговли был столь же неизбежен, сколько объективен и детерминирован. В основе этого процесса лежит резкое сокращение трансакционных издержек в электронной коммерции, поставившее эту форму экономической деятельности вне конкуренции (См. Табл. 1).

**Табл. 1. Сравнительная характеристика трансакционных издержек** [1, с. 32]

| Товарная группа | Себестоимость продаж в традиционных системах, US$ | Себестоимость электронных продаж, US$ |
|---|---|---|
| Программное обеспечение | 15,00 | 0,2-0,5 |
| Банковские услуги | 1,08 | 0,13 |
| Авиабилеты | 8,0 | 1,0 |
| Биллинг | 2,22-3,32 | 0,65-1,1 |
| Страховые полисы | 400-700 | 200-350 |
| Торговая наценка на продовольствие, % | 25-50 | 5-10 |

Не случайно Д.Норт определил «*институты как детерминирующие факторы экономического процесса, а изменения в соотношении цен – как источник институциональных изменений*» [6, с. 22]. Старые институты неэффективной структурой своих трансакционных издержек подтолкнули процесс формирования электронной коммерции, а сложившийся ценовой дисбаланс старого и нового стал источником её институционализации.

Происхождение институтов электронной коммерции вполне укладывается в рамки общих закономерностей любого институционального развития. Там, где созревают условия для возникновения новых форм экономической деятельности, рано или поздно возникают связанные с ними новые институты.

В неоинституциональной теории приводится два основных варианта возникновения экономических институтов [7, с. 161]:

1) *спонтанное* – в результате реализации индивидуальных устремлений участников рынка;

2) *преднамеренное* – в результате внешнего управляющего воздействия.

С другой стороны некоторые авторы предлагают использовать классификацию, подразумевающую разделение предпосылок формирования новых институтов по двум взаимоисключающим основаниям [8, с. 83-84]:

1) эндогенные предпосылки, основанные на принципе *методологического индивидуализма*, объясняющие институциональные процессы субъективными устремлениями экономических субъектов;

2) экзогенные предпосылки, основанные на принципе *методологического холизма*, объясняющие институциональные процессы их объективной востребованностью в обществе.

Определяющее влияние на процесс формирования новых институтов обычно приписывается эндогенным предпосылкам [9, с. 9-10]. Вместе с тем, о субъективном (индивидуалистическом, эндогенном, управляемом) характере институциональных процессов в электронной коммерции можно говорить, если эти процессы происходят в рамках традиционных экономических отношений традиционных экономических субъектов.

Институционализация электронной коммерции носит, наоборот, объективный (холистический, экзогенный, спонтанный) характер, так как субъектами новых отношений здесь являются абсолютно новые участники рынка. Практика свидетельствует о том, что ни одна из существовавших в традиционной экономике корпораций не смогла предложить ничего, что послужило бы толчком для институционального развития электронной коммерции.

Это всегда были новые экономические субъекты, начинавшие практически с нуля и в короткие сроки добивавшиеся впечатляющих результатов. Они спонтанно возникли в новом экономическом измерении на новых рынках под действием нового спроса, спровоцированного значительным снижением трансакционных издержек в виртуальной среде Интернета.

Если же обратиться к понятию «институциональный цикл», то кажущиеся противоречия в подходах снимаются [10, с. 90]. Так, спонтанное (экзогенное) формирование институтов присуще начальным этапам институционального цикла, а преднамеренное (эндогенное) – его завершающим этапам:

*Этап 1: Внедрение.* Спонтанно возникают новые методы экономической деятельности, основанные на использовании преимуществ от конкурентной минимизации трансакционных издержек. На этом этапе ещё нет формализованных институтов. Есть только более эффективная в сравнении с традиционными формами отношений экономическая практика. Причина её возникновения – экзогенна, т.к. обусловлена условиями внешней среды. Ещё К.Менгер писал о том, что в благоприятных экономических условиях субъекты рынка могут самоорганизовываться «*без какого-либо соглашения, без законодательного принуждения и даже без учёта общественных интересов*» [11, с. 164].

Институциональные формы, которые возникают на этом этапе, можно назвать институтами с большой натяжкой. В начале прошлого века немецкий экономист Г.Шмоллер трактовал такие институты как «*определённый порядок совместной жизни, который служит конкретным целям и обладает потенциалом самостоятельной эволюции*» [12, с. 61].

Уже на этом этапе могут формироваться робкие попытки нормативного закрепления институциональных рамок новых форм экономической деятельности. Например, в 1996 году Комиссией ООН по праву международной торговли (ЮНСИТРАЛ) был принят «Типовой закон ЮНСИТРАЛ об электронной торговле» [13]. Этот рекомендательный акт формулирует основные подходы к законодательному оформлению новых институтов.

Проблема была лишь в том, что спонтанная самоорганизация электронной коммерции ускоренно развивалась на уровне малого предпринимательства. В отличие от крупного бизнеса, малые предприниматели находятся зачастую вне сферы нормативного регулирования, не прибегая к использованию государственного арбитража и не платя налоги.

*Этап 2: Рост*. Новые методы экономической деятельности трансформируются в рутины, т.е. в общепринятые поведенческие алгоритмы, обеспечивающие сравнительную эффективность принятия решений. Применительно к электронной коммерции это означало формирование виртуального рынка со своими, присущими только ему правилами и нормами поведения.

Здесь можно уже говорить о завершении институционализации соответствующих рутин, выступающих в качестве информационной доминанты экономического поведения участников виртуального рынка. Указанная закономерность характерна не только для виртуального рынка, но и для любых рынков. «*Одной из важных функций институциализированных рутин,* – отмечал Дж.Ходжсон, – *является создание возможностей для снабжения информацией других агентов*» [14, с. 202].

На втором этапе институционализации электронной коммерции новые формы экономических отношений заметно теснят старые институты. Государство уже не может игнорировать меняющуюся экономическую реальность и постепенно начинает осознавать потребность в её нормативном регулировании. Электронная коммерция в России находится сегодня именно на этом этапе институционального развития.

*Этап 3: Зрелость*. На рынке электронной коммерции наблюдается укрупнение и специализация экономических субъектов, которые постепенно завоевывают и монополизируют рынок. Время энтузиастов заканчивается, уступая место времени профессионалов. Одновременно падает значение устаревших форм экономической деятельности: в конкурентных областях они поглощаются, трансформируются или просто уходят с рынка.

На уровне государства происходит признание экономической роли новых форм экономических отношений и их юридическая институционализация. Государство вводит юридические нормативы экономической деятельности и санкции за их несоблюдение. Институциональные процессы переходят в правовое русло. На смену предпринимательским структурам приходят корпорации.

В электронной коммерции этот процесс только начинается. В любом случае формирование нормативно-правовых институтов будет следствием формирования рутин и инфраструктуры электронной коммерции, а не наоборот [15, с. 35]. В России, например, даже государственная статистика пока в упор «не видит» и не отслеживает электронную коммерцию [16]. Для изменения ситуации требуется коренная ситуация не только форм государственной (налоговой, статистической и т.п.) отчётности, но и методов обработки макроэкономических данных.

*Этап 3: Упадок*. На рынке формируются новые формы экономических отношений, не укладывающиеся в традиционные институциональные рамки. Эффективность субъектов традиционной экономики падает, а их методы продвижения товаров устаревают. На смену устоявшимся экономическим отношениям и институтам приходят новые.

Сейчас трудно даже предположить, когда начнётся и с какими технологиями будет связан упадок сетевой экономики, основанной на методах электронной коммерции. Жизненный цикл институтов определяется жизненным циклом экономических отношений, а они могу десятилетиями и даже столетиями оставаться неизменными, хотя прогресс здесь, несомненно, ускоряется.

Гораздо важнее своевременно выявить, проанализировать и классифицировать происходящие структурные изменения в экономике. Затем на этой базе можно будет разрабатывать стратегии экономического роста, актуальные как на уровне отдельных субъектов, так и на уровне государственного управления. В эпоху информационной экономики информация определяет не только вектор экономического развития, но и его конкурентоспособность.

Лучшим критерием для проверки любой экономической теории выступает экономическая практика, которая свидетельствует о том, что электронная коммерция (и дропшиппинг в том числе) представляет собой процесс самостоятельного развития новых экономических отношений. Поэтому проблема институционализации электронной коммерции на данном этапе заключается не в столько в решении проблемы возникновения связанных с ней институтов, сколько в классификации её направлений, форм и методов. Практика здесь значительно опережает теорию, которая пока еще не сложилась в окончательном виде.

**Природа трансакционных издержек в электронной коммерции**. Важнейшую прикладную задачу неоинституциональной теории применительно к электронной коммерции можно определить как повышение эффективности торговых операций в заданных институциональных рамках. Речь идёт о теоретическом обосновании процессов достижения рыночными субъектами экономических целей в виртуальной среде Интернета. Пока такое обоснование не сформулировано, экономическая практика представляет собой набор не связанных между собой маркетинговых проектов с разной степенью успешности.

Решение проблемы может быть найдено на пути отказа от неоклассического постулата о нулевых трансакционных издержках, традиционно учитываемых, что называется, «при прочих равных» [9, с. 12]. Как уже говорилось выше, сетевая экономика (и неразрывно связанная с ней электронная коммерция) основана на перевёрнутой системе базовых ценностей. «При прочих равных» здесь учитываются производственные издержки, которые примерно сопоставимы у всех участников рынка в условиях глобальной мировой экономики.

Определяющие эффективность электронной коммерции трансакционные издержки сегодня не просто являются основополагающим понятием неоинституциональной теории. Их использование в научном обороте позволяет наилучшим образом объяснить происхождение феноменального успеха электронной коммерции в мировой и российской экономике.

Вместе с тем, трансакционные издержки в сетевой экономике имеют как минимум одно существенное отличие от аналогичных издержек в традиционной экономике. Классик неоинституциональной теории Д.Норт писал, что трансакционные издержки являются составной частью производственных издержек [6, с. 46]. В традиционной экономике это обстоятельство объясняется тем, что поставки контролируются производителями, которые являются неотъемлемой частью традиционных каналов распределения.

В сетевой экономике наблюдается обратная ситуация. В отличие от традиционной экономики, сетевая экономика представляет собой не экономику материального производства, а экономику виртуальных коммуникаций [17, с. 5]. Субъекты сетевых экономических отношений не производят материальные продукты. Их сфера деятельности – «производство» нематериальных коммуникативных услуг. Поэтому в сетевой экономике трансакционные издержки не могут являться частью производственных издержек в том смысле, в котором писал о них Д.Норт.

Дуглас Норт выделял два вида производственных издержек [6, с. 46]:

1) *трансформационные издержки*, связанные с изменением физических свойств материальных объектов;

2) *трансакционные издержки*, связанные с переходов прав собственности на товары.

Однако сетевая экономика не связана с физической трансформацией материальных объектов в производственном процессе. Мало того, она далеко не всегда связана с перехо-дом прав собственности. Например, в дропшиппинге не происходит реального перехода прав собственности, т.к. дропшипперы оперируют в сделках с поставщиками и покупателями не принадлежащими им ценностями.

В электронной коммерции мы вообще имеем дело с продажей информационных услуг, не связанных напрямую ни с производителем, ни с потребителем. Конечный продукт здесь представляет собой коммуникативные возможности, основанные на использовании интернет-технологий, выражающиеся в экономии сил, денег и времени заказчиков.

Не случайно определяющую роль в сокращении трансакционных издержек в сетевой экономике играет фактор информации. Здесь следует отметить и то, что некоторые неоинституционалисты, как например К.Менар, ещё до формирования сетевой экономики в её современном виде отмечали, что фактор информации определяет формализацию контрактных отношений [18, с. 75].

Неоинституциональная теория предполагает, что любое экономическое действие неразрывно связано с непроизводственными затратами. Поэтому в условиях глобальной экономики производственные издержки все больше переходят в категорию «при прочих равных», а конкуренция смещается в непроизводственную сферу. В результате рыночный потенциал снижения трансакционных издержек существенно превышает потенциал снижения производственных издержек и на первое место выходит фактор информации.

В условиях практически неограниченных коммуникативных и информационных возможностей сетевой экономики в конкурентной борьбе выигрывает тот, кто быстрее и с меньшими затратами принял, обработал и выполнил полученный заказ. Это необязательно должно быть связано с обработкой данных в Интернете и оказанием чисто информационных услуг.

Речь может идти, например, об оказании услуг транспортной логистики, связанных с доставкой товара покупателям. Условия поставок (цена, сроки, ответственность и т.д.) будут одинаковы в традиционной и в сетевой экономике. Однако приобретение услуг сетевого сервиса по обработке заказов будет подразумевать почти мгновенный приём заказов, их оплату и интерактивный контроль исполнения. Многократно увеличивается охват потенциальной аудитории, автоматизируются операции и резко снижается трудоёмкость обработки заказов. В результате бурный расцвет переживают как интернет-сервисы, оказывающие услуги транспортным компаниям, так и сами транспортные компании.

Аналогичным образом обстоит дело, например, с облачными сервисами, позволяющими перенести документооборот в виртуальную среду, сэкономив на программистах, серверах и оргтехнике [19]. Облачные сервисы не производят материальные продукты, а их услуги виртуальны. Однако именно эти платные услуги позволяют заказчикам выйти на новый уровень конкурентоспособности. Мало того, в условиях разделения прав собственности на средства производства (находящиеся, например, в Китае) и на технологии (например, из США или Европы), трансакционные преимущества гораздо важнее производственных преимуществ.

Поэтому **под трансакционными издержками в электронной коммерции предлагается подразумевать издержки экзогенного происхождения**, т.е. не зависящие от принимающего решения экономического субъекта. В отличие от эндогенных (внутренних) издержек, эти издержки определяются условиями рыночной (виртуальной, сетевой) среды. Здесь срабатывает своеобразный «закон вакуума», согласно которому преимущественное развитие наблюдается в тех секторах экономики, где происходит максимизация прибыли.

В условиях виртуальной маркетинговой среды особую актуальность получает тезис О.И.Уильямсона о том, что «*минимизация издержек более фундаментальна, чем разработка стратегий*» [20, с. 362]. В самом деле, экономический эффект от стратегического планирования довольно призрачен и зависит от множества неконтролируемых внешних факторов. Тогда как экономический эффект от минимизации трансакционных издержек проявляется немедленно и не зависит от колебаний рыночной конъюнктуры. Электронная коммерция позволяет преумножить эффект от снижения трансакционных издержек за счет многократного увеличения охвата целевой аудитории в Интернете.

Революция в сфере сокращения трансакционных издержек в электронной коммерции оказала огромное влияние на трансформацию маркетинга во Всемирной сети. Слабое место традиционного маркетинга всегда заключалось в преходящем характере маркетинговых достижений. Это связано с тем, что любое маркетинговое решение эффективно лишь до тех пор, пока его не приняли на вооружение конкуренты.

Итогом непрекращающейся маркетинговой гонки стало быстрое копирование и распространение любых сколько-нибудь эффективных маркетинговых новаций, что снижает и без того невысокую их эффективность. Не случайно в маркетинге последних десятилетий огромное значение приобрел бенчмаркинг – политика, основанная на своевременном копировании маркетинговых новаций наиболее успешных конкурентов [21].

С другой стороны, традиционные экономические субъекты до сих пор отягощены необходимостью поддержания существующей инфраструктуры сбыта и продвижения товаров. Тем самым они не только консервируют своё стратегическое отставание во внедрении новых форм коммерческой деятельности, но и создают благоприятные ценовые условия для их ускоренного развития. Кроме того, такая политика способствует непрекращающемуся оттоку покупателей в сетевую экономику.

В электронной коммерции, независимо от масштаба проводимых маркетинговых мероприятий, эффективность продаж в конечном итоге определяется эффективностью совершаемых покупок. Иначе говоря, чем выгоднее альтернативная сделка для покупателя, тем меньшую роль играют маркетинговые ухищрения продавца. Условия виртуального рынка в электронной коммерции наиболее приближены в этом смысле к условиям совершенной конкуренции.

Как и в традиционной экономике, покупателя здесь можно ввести в заблуждение относительно реального качества товара или его цены. Однако тактический успех вскоре неминуемо превратится в стратегическое поражение из-за утраты доверия покупателей. Особенность электронной коммерции заключается в глобальности Интернета: аналогичные по свойствам товары одновременно предлагает множество продавцов. При этом размер трансакционных издержек у продавцов и у покупателей несопоставимо ниже, чем в традиционной экономике, что нивелирует маркетинговое значение традиционного продвижения.

В виртуальном пространстве Интернета скорость неконтролируемого распространения информации и количество коммуникаций между потребителями возрастают многократно. Ни один продавец не может контролировать коммуникации во Всемирной сети, а потому реклама, ценообразование и другие традиционные инструменты маркетинга постепенно утрачивают свою актуальность. Любой покупатель за несколько минут может получить в Интернете максимум специальных знаний и о предлагаемом товаре и о его конкурентных аналогах.

При этом передача традиционно внутренних функций (логистики, коммуникаций и т.д.) внешним независимым посредникам привела к тому, что маркетинг обратил свой взор на проблематику экономии на издержках. В этой связи можно говорить о появлении в сетевой экономике новой разновидности маркетинга – ***трансакционного маркетинга***, направленного на повышение эффективности трансакций с независимыми партнёрами и покупателями внутри виртуальной сбытовой сети.

Маркетинговая деятельность, направленная на снижение трансакционных издержек, превратилась в электронной коммерции в важнейший инструмент конкурентной борьбы. В этой борьбе выигрывают участники рынка, максимизирующие свои маркетинговые возможности путём снижения издержек в результате расширенного использования возможностей интернет-коммуникаций.

В условиях мгновенности коммуникаций и безграничности виртуального пространства конкурентные преимущества можно получить только за счёт лучшей, чем у конку-рентов, логистической поддержки продаж. Именно этот механизм лежит в основе ускоренного развития подавляющего большинства форм и методов электронной коммерции.

**Библиографический список:**